\documentclass[twocolumn, prl, showpacs, preprintnumbers, amsmath,amssymb, superscriptaddress]{revtex4}

\usepackage{braket}

\pdfoutput=1

\usepackage{graphicx,epsf,epsfig,ulem}
\usepackage{epsfig}
\usepackage{epstopdf} 
	
\usepackage{bm}
\usepackage{subfigure}
\usepackage{color}
\usepackage{xcolor}

\begin{document}

\bibliographystyle{ieeetr}

\title{Two-electron states of a group V donor in silicon from atomistic full configuration interaction}
\textbf{}

\author{Archana Tankasala}
\affiliation{Network for Computational Nanotechnology, Purdue University, West Lafayette, IN 47907, USA}

\author{Joseph Salfi}
\affiliation{Centre for Quantum Computation and Communication Technology, School of Physics, The University of New South Wales, Sydney, 2052 New South Wales, Australia}

\author{Juanita Bocquel}
\affiliation{Centre for Quantum Computation and Communication Technology, School of Physics, The University of New South Wales, Sydney, 2052 New South Wales, Australia}

\author{Benoit Voisin}
\affiliation{Centre for Quantum Computation and Communication Technology, School of Physics, The University of New South Wales, Sydney, 2052 New South Wales, Australia}

\author{Muhammad Usman}
\affiliation{Centre for Quantum Computation and Communication Technology, School of Physics, The University of Melbourne, Parkville, 3010 Victoria, Australia}

\author{Gerhard Klimeck}
\affiliation{Network for Computational Nanotechnology, Purdue University, West Lafayette, IN 47907, USA}

\author{Michelle Y. Simmons}
\affiliation{Centre for Quantum Computation and Communication Technology, School of Physics, The University of New South Wales, Sydney, 2052 New South Wales, Australia}

\author{Lloyd C. L. Hollenberg}
\affiliation{Centre for Quantum Computation and Communication Technology, School of Physics, The University of Melbourne, Parkville, 3010 Victoria, Australia}

\author{Sven Rogge}
\affiliation{Centre for Quantum Computation and Communication Technology, School of Physics, The University of New South Wales, Sydney, 2052 New South Wales, Australia}

\author{Rajib Rahman}
\affiliation{Network for Computational Nanotechnology, Purdue University, West Lafayette, IN 47907, USA}

\date{\today}

\begin{abstract}
Two-electron states bound to donors in silicon are important for both two qubit gates and spin readout. We present a full configuration interaction technique in the atomistic tight-binding basis to capture multi-electron exchange and correlation effects taking into account the full bandstructure of silicon and the atomic scale granularity of a nanoscale device. Excited $s$-like states of $A_1$-symmetry are found to strongly influence the charging energy of a negative donor centre. We apply the technique on sub-surface dopants subjected to gate electric fields, and show that bound triplet states appear in the spectrum as a result of decreased charging energy. The exchange energy, obtained for the two-electron states in various confinement regimes, may enable engineering electrical control of spins in donor-dot hybrid qubits.
\end{abstract}


\maketitle 

\section{Introduction}
Quantum computing device architectures based on donors in silicon have attracted considerable attention in recent times \cite{kane1998silicon}. Electrons bound to phosphorous donors in silicon have long decoherence times \cite{tyryshkin2012electron,veldhorst2014addressable} due to weak spin-orbit coupling and small fraction of spin carrying isotopes of silicon. In addition to the well-studied neutral donor state (D$^0$) of a Group V donor in silicon, the two-electron negatively charged donor state (D$^-$) is of technological relevance \cite{kane2002electron,hollenberg2004single,greentree2005electrical,sellier2006transport,pierre2010single,lansbergen2011lifetime,watson2015high,harvey2015nuclear,wang2016highly,prati2016band}. 

The two-electron states of donors are important for spin readout through spin-to-charge conversion \cite{kane2002electron,hollenberg2004single,greentree2005electrical} and spin dependent tunneling \cite{watson2015high} and also for tuning the exchange coupling in two-qubits towards the charge transfer regime \cite{wang2016highly}. Recent experiments have also used the two-electron donor state with a bound hole for addressing nuclear spins \cite{steger2012quantum,saeedi2013room}. Moreover, the two-electron donor states are also observed in quantum transport in extremely-scaled field-effect transistors \cite{sellier2006transport,pierre2010single,lansbergen2011lifetime} and in artificially patterned dopant arrays that provide access to impurity Hubbard bands \cite{prati2016band}. Recent schemes to form hybrid donor-dot qubit systems that pulse electrons between donors and interface confined states also benefit from an understanding of the D$^-$ in a gated nanoscale environment \cite{harvey2015nuclear}.  
 
Two-electron states of donors in silicon, both in bulk and close to the oxide-semiconductor interfaces, have been previously studied from a mean field self-consistent tight-binding \cite{rahman2011efieldDm} and also variationally with Chandrasekhar type wavefunctions \cite{calderon2010heterointerface} to obtain ground states and charging energies. However, these approximate methods consider only Coulomb repulsion between electrons or treat higher order exchange-correlations only approximately. Consequently, the excited states of a two-electron system and the singlet-triplet splitting of relevance to qubits cannot be evaluated accurately from these models. In this work, we compute the two-electron states of the donor in a device-like environment from an atomistic full configuration interaction (FCI) method, which also provides excited states and exchange energies as a function of gate fields and donor locations. Moreover, we are able to evaluate the contribution of the various donor orbital and valley states to the two-electron state and show how the charging and exchange energies change with applied electric fields and donor depths in a device. With increasing electric field, we see the exchange changes non-monotonically for the donor-hybrid-interface sequence with applied fields. The results show how exchange energy between electrons confined in donors and interface states can be tuned in a hybrid donor-dot setting. 

The electronic structure of a group V donor in silicon is complicated by the six-fold conduction band valley degeneracy of silicon. The tetrahedral symmetry of the crystal and central-cell corrections \cite{kohn1955theory} lift the valley degeneracies and cause donor states to form hybrid valley-orbit states with species dependent level splittings \cite{ramdas1981spectroscopy}. We employ a multi-million atom tight-binding model to capture the peculiar electronic structure of the donor which is also affected by fields and interfaces present in a nanoscale device \cite{rahman2009orbital}. We use these donor states as basis functions to solve the two-electron problem from FCI, which provides converged two-electron states and their configurations in terms of Slater determinants \cite{szabo2012modern}. In this way, we include atomistic details of the silicon crystal and the nano-device into a description of exchange and correlation energies. This enables us to analyze the two-electron donor states in great detail and to gain insight on how to engineer them for single atom electronics in silicon.

\section{Methodology}

Single electron states of a single donor are obtained from the atomistic tight-binding (TB) approach \cite{slater1954wave}. The method is shown to correctly determine the D$^0$ binding energies of the donor and its excited states \cite{rahman2007high}. These are used as a basis to construct the many-electron anti-symmetric Slater determinants. In FCI, a multi-electron state is a superposition of different Slater configurations. Excited configurations arising from exciting electrons to virtual orbitals are considered to include higher order correlations. The multiply-excited Slater determinants form a complete basis set in the function space spanned by the single-electron states. FCI wavefunction for an $n$-electron system formed from a basis of $N$ single-electron states is expressed as:
\begin{gather*} 
\Ket{\Psi(\vec{r}_{1},..,\vec{r}_{n})}=C_{0}\Ket{\Psi_{0}}+\sum\limits_{a,r}C_{a}^{r}\Ket{\Psi_{a}^{r}}+\sum\limits_{ab,rs}C_{ab}^{rs}\Ket{\Psi_{ab}^{rs}}+...
\end{gather*} 
where, $\Ket{\Psi_{0}}$ is the ground state Slater determinant from $n$ occupied orbitals $\Ket{ab...c}$, $\Ket{\Psi_{a}^{r}}$ are singly-excited Slater determinants from promoting electron in state $\psi_{a}$ to excited state $\psi_{r}$, $\Ket{rb...c}$, $\Ket{\Psi_{ab}^{rs}}$ are doubly-excited Slater determinants from promoting electrons in states $\psi_{a}$ and $\psi_{b}$ to excited states $\psi_{r}$ and $\psi_{s}$, and so on\cite{szabo2012modern}. The $C_i$s are the corresponding coefficients of these Slater determinants. The multi-electron state, expanded in this basis, gives a more complete quantitative description of the configurations that contribute to the many-electron states.

In the basis of possible $n$-electron Slater determinants $\Ket{A}=\Ket{\psi_{i}\psi_{j}...}$ and $\Ket{B}=\Ket{\psi_{k}\psi_{l}...}$, the Hamiltonian element ${\cal H }(a,b)$ is $\Bra{\psi_{i}\psi_{j}...}{\cal H }\Ket{\psi_{k}\psi_{l}...}$ where the $\psi_{x}$ are the single-electron states of the system, in the basis of localized atomic orbitals of each atom in the crystal, as obtained from atomistic TB method. Electron-electron interactions in the Hamiltonian are of the form 
\begin{gather*}
\Bra{\psi_{i}\psi_{j}}\frac{e^2}{4\pi\epsilon|\vec{r}_{1}-\vec{r}_{2}|}\Ket{\psi_{k}\psi_{l}}
\end{gather*}
that need an evaluation of Coulomb and exchange integrals,
\begin{gather*}
J = \iint_{V} \! \psi_{i}^{*}(\vec{r}_{1})\psi_{j}^{*}(\vec{r}_{2})\frac{e^2}{4\pi\epsilon|\vec{r}_{1}-\vec{r}_{2}|}\psi_{k}(\vec{r}_{1})\psi_{l}(\vec{r}_{2}) \, \mathrm{d}\vec{r}_{1} \mathrm{d}\vec{r}_{2} \\
K = \iint_{V} \! \psi_{i}^{*}(\vec{r}_{1})\psi_{j}^{*}(\vec{r}_{2})\frac{e^2}{4\pi\epsilon|\vec{r}_{1}-\vec{r}_{2}|}\psi_{l}(\vec{r}_{1})\psi_{k}(\vec{r}_{2}) \, \mathrm{d}\vec{r}_{1} \mathrm{d}\vec{r}_{2}
\end{gather*}
respectively, for each Hamitonian element. Here, $\vec{r}_{1}$ and $\vec{r}_{2}$ are the coordinates of the two electrons, $V$ is the simulation domain, $e$ is the electronic charge and $\epsilon$ the dielectric constant of the host material.
 
Each of J and K is a double integral over the entire region of simulation domain. The evaluation of these integrals and the construction of an atomistic FCI Hamiltonian is computationally more expensive than the diagonalization of the FCI Hamiltonian to obtain multi-electron energies and wavefunctions. The computation is massively parallelized for large systems with many single-electron states in the basis. The integrals, defining an $n$-body problem, are evaluated using the Fast Multipole Method (FMM) \cite{yokota2011treecode} to reduce the computational complexity, when a high accuracy in the energies is not required. For a high accuracy in the solution, in the scale of $\mu eV$, FMM and brute force calculations have the same computational burden. Solving the FCI Hamiltonian gives the energies of the multi-electron states and the contributions from the possible Slater configurations to each of these states.

The binding energy D$^-_{BE}$ and charging energy D$^-_{CE}$ of the two-electron state are calculated from the two-electron ground state D$^-_{GS-FCI}$ obtained from atomistic FCI, the conduction band minima of silicon CB$_{min}$ and the single-electron ground state D$^0_{GS-TB}$ obtained from tight-binding simulations using the equations
\begin{align*}
 D^-_{BE}&=D^-_{GS-FCI}-CB_{min}-D^0_{GS-TB} \\
 D^-_{CE}&=D^-_{GS-FCI}-2D^0_{GS-TB}
\end{align*}

For donors located close to a Si-SiO$_2$ interface in a metal-oxide-semiconductor (MOS) device, the electrostatics of the metal-insulator-silicon considerably influences the energy spectrum of the donor, especially for a negative donor. The dielectric mismatches renormalize the electron-electron interaction. This effect can be captured in the atomistic FCI using the method of images charges. Here, the Coulomb and exchange interactions of the electron are evaluated, not only between two electrons, but also between an electron and the images of the other electron. Benchmarking the potential profile of a point-charge from the analytical method of image charges with COMSOL simulation software suggests that at least 10 images of each electron must be considered in the analytical method to obtain the correct potential. This considerably increases the computational burden in evaluation of each matrix element of FCI Hamiltonian. However, the inclusion of these image charges, to capture the screening of electron interactions, is necessary for an accurate quantitative evaluation of the two-electron states of the sub-surface donor. A similar method, tight-binding with FCI, used fast Fourier transforms to solve for the two-electron integrals and applied the method to study two-electron quantum dot states \cite{nielsen2012many}. 

Interface donor systems have been previously studied for arsenic donors in silicon using self-consistent mean-field approach\cite{rahman2011efieldDm}. This method is shown to give a good estimate of the D$^-$ binding energies in 10-20 iterations, but exchange correlations are ignored and the excited two-electron states and their configurations may not be exactly determined from this method. The Hartree method self-consistently updates the basis in several iterations. However, in general, FCI does not scale well with the number of electrons.

\section{Results and Analysis}

The electronic spectrum of a single donor in a tetrahedrally symmetric crystal potential of silicon is modified by valley-orbit coupling. Here the six-fold valley degeneracy of the ground state, the $1s$ manifold, of donor Coulomb confinement, is broken by the local crystal field into a singlet with $A_1$ symmetry, a triplet with $T_2$ symmetry and a doublet with $E$ symmetry as shown in Fig.1(a). The binding energies and degeneracies of these $1s$ states of a phosphorous donor in silicon, D$^0$ center, have been determined using semi-empirical tight-binding model and successfully validated against experimental values \cite{rahman2007high,salfi2014spatially}. For donors in the vicinity of a gate, the Stark shifted donor spectrum calculated using the tight-binding method also agrees closely with the experiments \cite{lansbergen2008gate}.

In this work, D$^-$ centers are studied using atomsitic FCI and the two-electron charging energies of bulk and interface donors are compared against available experimental values. Excited states of the donor spectrum $2p$, $2s$...up to $4s$ and their degeneracies are also resolved from the eigensolver, which are crucial to solve the two-electron problem as shown later. 

\subsection{Negative donor in bulk}

Optical experiments on a bulk phosphorous donor in silicon indicate the presence of a bound singlet D$^-$ state with a binding energy approximately 2.0meV below the conduction band minimum, corresponding to a charging energy of 43.6meV \cite{taniguchi1976d}. Fig. 1 presents atomistic FCI calculation of a bulk phosphorous donor D$^-$ in silicon. It must be noted that although we treat phosphorous donor specifically here, the same method can be used on other shallow group V donors in silicon and the results will qualitatively be the same. Fig. 1(b) shows the charging energy of the two-electron state as a function of the number of spin-resolved D$^0$ atomic states in the basis. Including only $1s$-like D$^0$ states in the atomistic FCI method gives a significantly overestimated charging energy of 71.8meV as shown in Fig. 1(b), indicating an incomplete basis. With the inclusion of $2s$, $3s$ and $4s$-like D$^0$ states in the basis, the charging energy converges to 46.0meV. It must be noted that the binding energies are evaluated with reference to the conduction band minima of bulk silicon at 1.131355eV obtained from a bulk bandstructure calculation \cite{boykin2004valence}. A finite box size of the simulation domain, in this case 35nm x 35nm x 35nm causes interface confinement effects on the excited donor states. Therefore, the system is not really bulk, only bulk-like, which leads to higher orbital energies due to increased confinement. This is one of the reasons for slightly higher charging energy of 46.0meV relative to the 43.6meV measured in the optical experiments. From the wavefunction, the D$^-$ state with 46.0meV charging energy is observed to be localized to the donor. Within the approximations of tight-binding basis orbitals, discretized tight-binding wavefunctions, central-cell corrected donor potential and truncated single-electron basis set, atomistic FCI still gives a fairly close charging energy of the negative phosphorous donor in bulk.  Hence, the excited donor states in the manifold of $2p$, $2s$, $3s$ and so on, which are typically ignored in most calculations, are crucial for the solution of the two-electron donor state. 

\begin{figure}[htbp]
\center\epsfxsize=3.4in\epsfbox{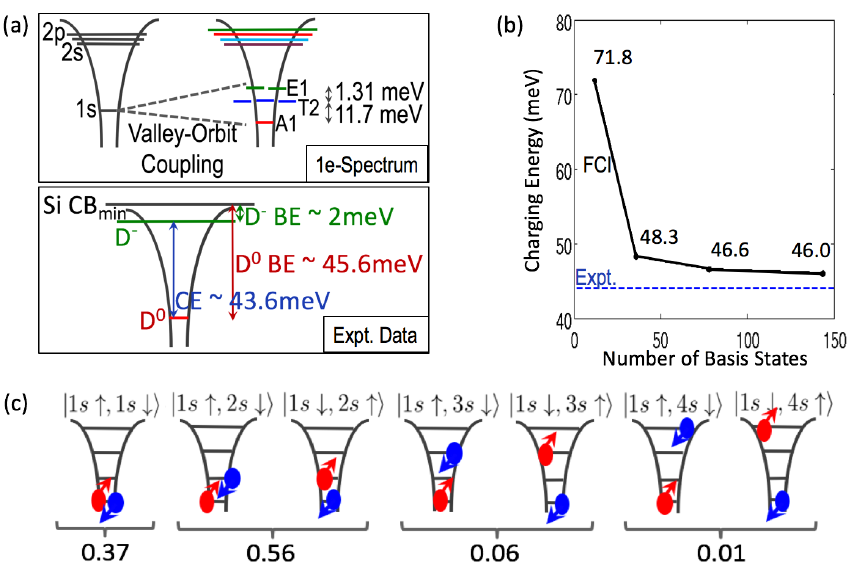}
\caption{Two-electron state of a bulk P donor in silicon; (a) Degeneracy of hydrogenic $1s$-like states of a donor Coulomb confinement potential, broken into valley-orbit singlet, triplets and doublets ($A_1$, $T_2$ and $E$ respectively) by the tetrahedral crystal symmetry of the host material. Binding and charging energies of donor bound electrons from experimentally observed $D^0$ and $D^-$ states. (b) $D^-$ charging energy of phosphorous donor in silicon converges towards the experimentally observed value of 43.6meV \cite{taniguchi1976d} with increasing number of single electron states in the FCI basis. The data points correspond to the inclusion of $1s$, $2s$, $3s$ and $4s$-like states in the basis. (c) Two-electron $D^-$ ground state. Numbers indicate the probability that the system is in each of the configurations. One of the electrons is always in the $D^0$ ground state, $1s$-like state with $A_1$ symmetry. The second electron occupies $2s$, $3s$ or $4s$-like excited states also of $A_1$ symmetry. The excited state splittings corresponding to 2s, 3s, 4s are magnified in the schematic relative to Fig. 1(a) to portray the electronic configurations clearly. Also, the $T_2$ and $E_1$ states are omitted for simplicity.}
\vspace{0cm}
\label{fi6}
\end{figure}

Fig. 1(c) shows the two-electron configurations that compose the singlet D$^-$ ground state of a phosphorous donor in silicon as obtained from FCI. From the two-electron FCI wavefunction, it is seen that the probability of the D$^-$ state existing in $\Ket{1s\uparrow,1s\downarrow}$ configuration is only 37\%. $1s$-like states are therefore not sufficient to describe the D$^-$ ground state. There is a significant contribution of 56\% from $\Ket{1s\uparrow,2s\downarrow}$ and $\Ket{1s\downarrow,2s\uparrow}$ states and including the $2s$-like states in the FCI basis corrects the charging energy. Contributions from $2s$, $3s$ and $4s$-like D$^0$ states to the D$^-$ ground state suggests that the negative donor has strong Coulomb correlations. A complete basis for such a system is huge and requires at least 150 single electron spin states. For the employed basis and simulation domain, the calculated atomistic FCI charging energy (46.0 meV) approaches the value in bulk (43.6 meV).  The remaining discrepancy is small and might be reduced by increasing basis size or simulation domain size. However, increasing either is beyond the scope of the present work.

As can be seen from Fig. 1(c), all the contributing electronic configurations have anti-parallel spins, consistent with the net spin of a singlet state. Moreover, we find that one of the electrons of D$^-$ always occupies the $1s$-like state with $A_1$ symmetry, the D$^0$ ground state. It is favorable for the other electron to have the same valley symmetry which increases the electron correlations, thus lowering the energy of the two-electron singlet state. From the FCI solution of the ground state, we find that $2s$, $3s$ and $4s$-like D$^0$ states that contribute to the D$^-$ singlet have all $A_1$ symmetry.  Therefore, for the bulk donor, $T_2$ and $E$ states with different valley symmetries are not found to contribute to the D$^-$ ground state. It must be noted that the D$^-$ state from atomistic FCI is somewhat like the Chandrasekhar-like wavefunction with appropriate valley symmetries, if the second orbital of the Chandrasekhar wavefunction can be thought of as a hybrid of the $1s$, $2s$, $3s$ and $4s$ orbitals.

The significant contributions of the excited Slater determinants to the D$^-$ state show strong electronic correlations based on orbital and valley symmetries. Such correlations cannot be captured with a mean-field treatment of the problem. Atomistic FCI, therefore, provides a comprehensive description and valuable insights into the two-electron states of donors in silicon taking into account the electron-electron interactions arising from excited configurations that cannot be ignored for a correlated system.  

\subsection{Negative donor close to interface} 

Donors close to the oxide-silicon interface are important for silicon quantum computing for a number of reasons. The first signature of single donor orbital states and their Stark shift were detected in single donors located less than 10nm from an oxide-silicon interface in FinFETs \cite{lansbergen2011lifetime}. In this regime, the donor is strongly tunnel coupled to states confined at the interface, and a gate voltage can ionize the electron adiabatically through intermediate donor-interface/donor-dot hybrid states \cite{calderon2006quantum,rahman2011efieldDm}. There have been proposals to use this system to form hybrid donor dot qubits where electrons are pulsed to the interface states for two-qubit operations with enhanced exchange couplings and pulsed back to the donors to take advantage of the long quantum memory in donor bound states \cite{calderon2009quantum,pica2016surface}. Recent experiments have also manipulated multiple electrons between donors and interface states \cite{harvey2015nuclear} in a hybrid donor-dot setting. 

\begin{figure}[htbp]
\center\epsfxsize=3.4in\epsfbox{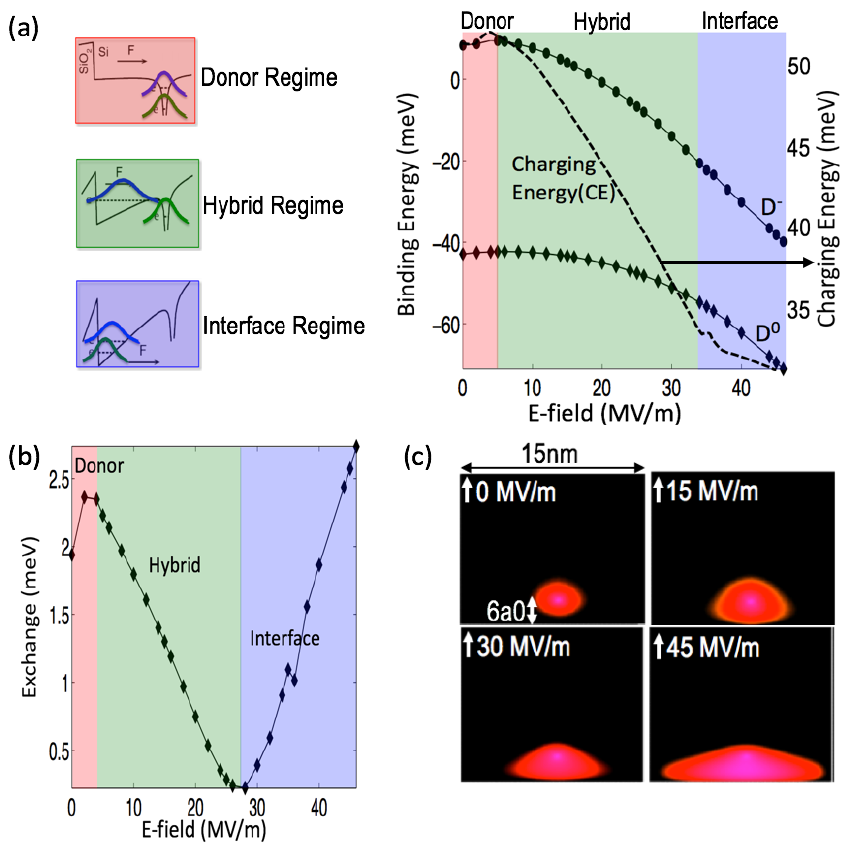}
\caption{Effect of electric field on the charging and exchange energies of a P donor in silicon, 6a0 deep from oxide-semiconductor interface; (a) Schematic of possible regimes with increasing E-fields when both electrons are localized at donor, one electron at donor and other hybridized with interface, both electrons at the interface. D$^0$ and D$^-$ binding energies and derived D$^-$ charging energy (dashed line) with increasing electric fields. (b) Exchange or singlet-triplet splitting of the D$^-$ state. (c) Two-electron densities of D$^-$ singlet ground state from full configuration interaction for different confinement regimes (on a $log$ scale). Delocalization of two-electron states is observed at large fields.}
\vspace{0cm}
\label{fi3}
\end{figure}

Two-electron charging energies of sub-surface dopants have also been observed to be around 30meV \cite{lansbergen2011lifetime} in stark contrast to bulk charging energies close to 44meV. This also enables the possibility of bound triplet states, which to our knowledge have not been observed in bulk donors, and a measurable singlet-triplet (exchange) splitting. Here, we show FCI can explain these experimental observations. We will also show the change in exchange coupling in a two-electron donor state as a function of an applied bias for donors close to interfaces, which is of relevance in hybrid donor-dot qubits.

The binding energies of both D$^0$ and D$^-$ states have been plotted in Fig. 2(a). The difference between the two is the charging energy of D$^-$, shown by a dashed line on the right axis. Singlet-triplet splitting or the exchange energy, with varying electric field is shown in Fig. 2(b). Exchange depends on the overlap of the single electron wavefunctions. Charge densities of the D$^-$ ground state are shown in Fig. 2(c) revealing the spatial spread of two-electron states under different confinement regimes.

Increasing the electric field causes the excited donor states to hybridize with the interface well states. One of the electrons is pushed towards the interface leading to delocalization of the electron, as shown in the schematic for hybrid regime in Fig. 2(a). With increasing field in this regime (at moderate electric fields) the delocalization increases which lowers the charging energy. This delocalization also lowers the overlap of the electron wavefunctions, thus decreasing the exchange with increasing fields.

At large fields, both electrons dominantly occupy the ground state of the interface well, strongly confined along the vertical field direction and weakly confined by the donor potential laterally. The decrease in orbital energies of single-electron states with increasing field leads to the decrease in the charging energy of D$^-$ in the interfacial regime. However, the increase in wavefunction overlap of the two electrons localized at the interface causes an increase in exchange with increasing electric field. The donor regime, at low electric fields, is similar to the interface regime, where both electrons are localized at the donor instead of the interface.  

\begin{figure}[htbp]
\center\epsfxsize=3.4in\epsfbox{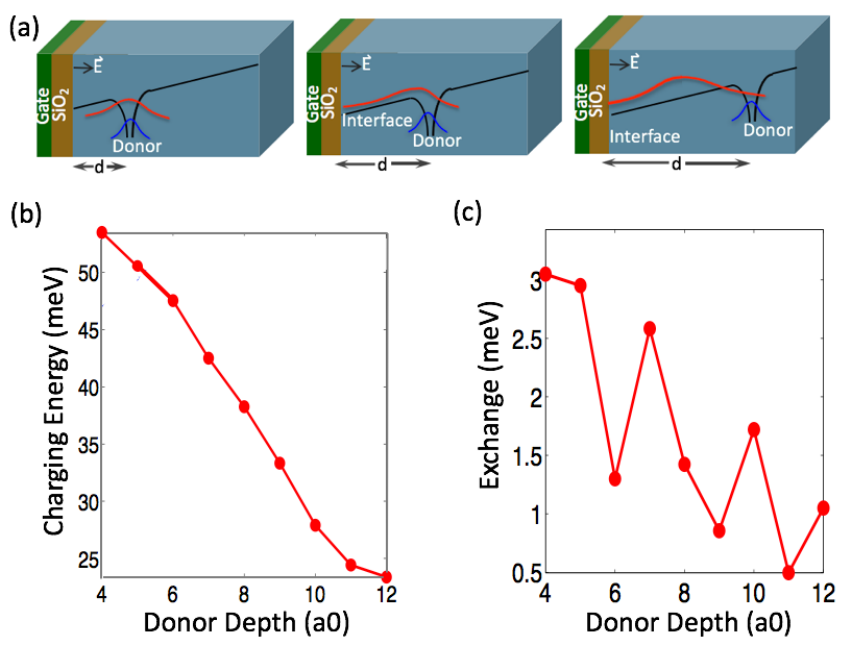}
\caption{Effect of the donor depth on the charging and exchange energies; (a) Schematic of increasing electron delocalization with increasing donor depth at a moderate electric field of 15MV/m, in the hybrid regime. (b) D$^-$ charging energies. (c) Singlet-triplet splitting of D$^-$. (a0=0.543nm for silicon)}
\vspace{0cm}
\label{fi4}
\end{figure}

D$^-$ charging and exchange energies varying with depth of the P donor are shown in Fig. 3 corresponding to the hybrid regime. At 15MV/m field, as can be understood from the schematic in Fig. 3(a), increasing the depth of the donor increases the delocalization of one of the electrons, thus lowering the charging energies as seen in Fig. 3(b). This also leads to a smaller overlap of electron wavefunctions and therefore, exchange also decreases with increasing donor depth as shown in Fig. 3(c). The oscillations in exchange are due to the difference in phase of the Bloch wavefunctions of the two electrons.

\begin{figure}[htbp]
\center\epsfxsize=3.4in\epsfbox{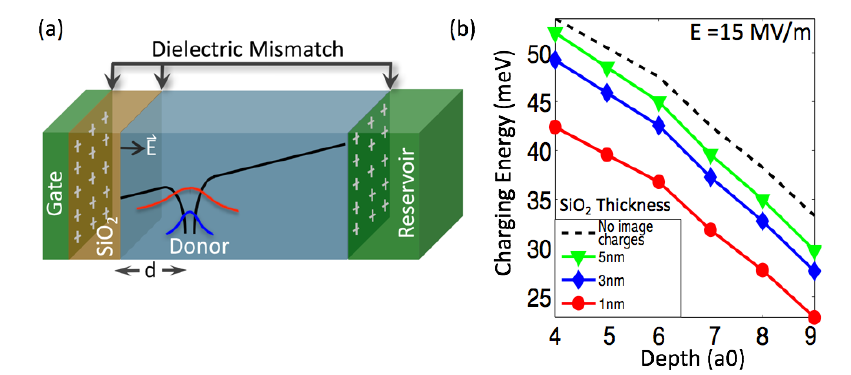}
\caption{Hetero-interface effects on a donor charging energy; (a) P donor in silicon, located close to Si-SiO$_2$-metal interface and to a heavily doped reservoir, important in resonant tunneling experiments. (b) Charging energy of D$^-$ at 15MV/m E-field for different oxide thicknesses. A thinner dielectric increases the screening of electronic repulsions by the metallic gate, leading to lower charging energies.}
\vspace{0cm}
\label{fi4}
\end{figure}

Donors close to the interface have dielectric mismatches arising from Si-SiO$_2$ and SiO$_2$-metallic gate interfaces. A reservoir of heavily doped silicon substrate also creates a dielectric mismatch as shown in Fig. 4(a). The reservoir is particularly important in scanning tunneling experiments when the highly doped silicon layer is located 10-20nm away from the donor \cite{voisin2015spatially,salfi2017robust}. For MOS-like devices, the device contacts are in proximity to the donor too and ideally the image charges in contacts must be accounted for as well, which have been ignored in this work \cite{voisin2014control}. Metallic and oxide interfaces close to an electron affect the electrostatics of the system and screen the electron-electron repulsions, thus lowering the two electron energies of the system. Fig. 4(b) shows the effect of the electrostatics of interfaces on the charging energy of D$^-$ state. As the thickness of the SiO$_2$ oxide layer decreases, the Si-SiO$_2$-metal interface becomes more metallic in nature and screens the electron interactions resulting in decreasing the two electron charging energy. The effect of the image charges is seen mainly as an offset in the charging energy. 

\section{Conclusion}

Atomistic FCI simulations demonstrate that the bulk D$^-$ is a highly correlated state with significant contributions from higher $s$-like D$^0$ orbitals of $A_1$ symmetry to the two-electron singlet state. Charging energy of a bulk negative donor is significantly overestimated without inclusion of these excited levels. For shallow donors, the charging and exchange energies are shown to be sensitive to the depth of the donors from the Si-SiO$_2$ interface, applied electric fields and the electrostatics of the interfaces, from the atomistic FCI simulations. The charging and exchange energies of D$^-$ are lower for deeper donors under moderate electric fields, typically the regime of interest for hybrid donor-dot architectures, but the exchange energies show small oscillations, less than an order of magnitude, with increasing depth. Moreover, in spite of the increased computational complexity in including the effect of electrostatics of interfaces on electron interactions, they are significant in evaluating the charging energies of shallow donors in the vicinity of oxide and metallic interfaces. The understanding of two-electron states from this work may be useful in the realization of hybrid donor-dot qubit architectures. \\
     
\begin{acknowledgments}
This work is funded by the ARC Center of Excellence for Quantum Computation and Communication Technology (CE1100001027), and in part by the U.S. Army Research Office (W911NF-08-1-0527). MYS acknowledges an ARC Laureate Fellowship. Computational
resources are acknowledged from NCN/Nanohub. This work is also part of the Accelerating Nano-scale Transistor Innovation with NEMO5 on Blue Waters PRAC allocation support by the National Science Foundation (award number OCI-0832623).
\end{acknowledgments}

Electronic address: atankas@purdue.edu, rrahman@purdue.edu

\vspace{-0.5cm}

\bibliography{references}

\begin{thebibliography}{10}

\bibitem{kane1998silicon}
B.~E. Kane, ``A silicon-based nuclear spin quantum computer,'' {\em Nature},
  vol.~393, no.~6681, pp.~133--137, 1998.

\bibitem{tyryshkin2012electron}
A.~M. Tyryshkin, S.~Tojo, J.~J. Morton, H.~Riemann, N.~V. Abrosimov, P.~Becker,
  H.-J. Pohl, T.~Schenkel, M.~L. Thewalt, K.~M. Itoh, {\em et~al.}, ``Electron
  spin coherence exceeding seconds in high-purity silicon,'' {\em Nature
  Materials}, vol.~11, no.~2, pp.~143--147, 2012.

\bibitem{veldhorst2014addressable}
M.~Veldhorst, J.~Hwang, C.~Yang, A.~Leenstra, B.~De~Ronde, J.~Dehollain,
  J.~Muhonen, F.~Hudson, K.~M. Itoh, A.~Morello, {\em et~al.}, ``An addressable
  quantum dot qubit with fault-tolerant control-fidelity,'' {\em Nature
  Nanotechnology}, vol.~9, no.~12, pp.~981--985, 2014.

\bibitem{kane2002electron}
B.~Kane, ``Electron devices for single electron and nuclear spin measurement,''
  Apr.~9 2002.
\newblock US Patent 6,369,404.

\bibitem{hollenberg2004single}
L.~Hollenberg, C.~Wellard, C.~I. Pakes, and A.~Fowler, ``Single-spin readout
  for buried dopant semiconductor qubits,'' {\em Physical Review B}, vol.~69,
  no.~23, p.~233301, 2004.

\bibitem{greentree2005electrical}
A.~D. Greentree, A.~Hamilton, L.~C. Hollenberg, and R.~Clark, ``Electrical
  readout of a spin qubit without double occupancy,'' {\em Physical Review B},
  vol.~71, no.~11, p.~113310, 2005.

\bibitem{sellier2006transport}
H.~Sellier, G.~Lansbergen, J.~Caro, S.~Rogge, N.~Collaert, I.~Ferain,
  M.~Jurczak, and S.~Biesemans, ``Transport spectroscopy of a single dopant in
  a gated silicon nanowire,'' {\em Physical Review Letters}, vol.~97, no.~20,
  p.~206805, 2006.

\bibitem{pierre2010single}
M.~Pierre, R.~Wacquez, X.~Jehl, M.~Sanquer, M.~Vinet, and O.~Cueto,
  ``Single-donor ionization energies in a nanoscale cmos channel,'' {\em Nature
  Nanotechnology}, vol.~5, no.~2, pp.~133--137, 2010.

\bibitem{lansbergen2011lifetime}
G.~Lansbergen, R.~Rahman, J.~Verduijn, G.~Tettamanzi, N.~Collaert,
  S.~Biesemans, G.~Klimeck, L.~Hollenberg, and S.~Rogge, ``Lifetime-enhanced
  transport in silicon due to spin and valley blockade,'' {\em Physical Review
  Letters}, vol.~107, no.~13, p.~136602, 2011.

\bibitem{watson2015high}
T.~Watson, B.~Weber, M.~House, H.~B{\"u}ch, and M.~Simmons, ``High-fidelity
  rapid initialization and read-out of an electron spin via the single donor
  {D}- charge state,'' {\em Physical Review Letters}, vol.~115, no.~16,
  p.~166806, 2015.

\bibitem{harvey2015nuclear}
P.~Harvey-Collard, N.~T. Jacobson, M.~Rudolph, J.~Dominguez, G.~A.~T. Eyck,
  J.~R. Wendt, T.~Pluym, J.~K. Gamble, M.~P. Lilly, M.~Pioro-Ladri{\`e}re, {\em
  et~al.}, ``Nuclear-driven electron spin rotations in a single donor coupled
  to a silicon quantum dot,'' {\em arXiv preprint arXiv:1512.01606}, 2015.

\bibitem{wang2016highly}
Y.~Wang, A.~Tankasala, L.~C. Hollenberg, G.~Klimeck, M.~Y. Simmons, and
  R.~Rahman, ``Highly tunable exchange in donor qubits in silicon,'' {\em NPJ
  Quantum Information}, vol.~2, p.~16008, 2016.

\bibitem{prati2016band}
E.~Prati, K.~Kumagai, M.~Hori, and T.~Shinada, ``Band transport across a chain
  of dopant sites in silicon over micron distances and high temperatures,''
  {\em Scientific Reports}, vol.~6, 2016.

\bibitem{steger2012quantum}
M.~Steger, K.~Saeedi, M.~Thewalt, J.~Morton, H.~Riemann, N.~Abrosimov,
  P.~Becker, and H.-J. Pohl, ``Quantum information storage for over 180s using
  donor spins in a 28{S}i semiconductor vacuum,'' {\em Science}, vol.~336,
  no.~6086, pp.~1280--1283, 2012.

\bibitem{saeedi2013room}
K.~Saeedi, S.~Simmons, J.~Z. Salvail, P.~Dluhy, H.~Riemann, N.~V. Abrosimov,
  P.~Becker, H.-J. Pohl, J.~J. Morton, and M.~L. Thewalt, ``Room-temperature
  quantum bit storage exceeding 39 minutes using ionized donors in
  silicon-28,'' {\em Science}, vol.~342, no.~6160, pp.~830--833, 2013.

\bibitem{rahman2011efieldDm}
R.~Rahman, G.~Lansbergen, J.~Verduijn, G.~Tettamanzi, S.~Park, N.~Collaert,
  S.~Biesemans, G.~Klimeck, L.~Hollenberg, and S.~Rogge, ``Electric field
  reduced charging energies and two-electron bound excited states of single
  donors in silicon,'' {\em Physical Review B}, vol.~84, no.~11, p.~115428,
  2011.

\bibitem{calderon2010heterointerface}
M.~Calder{\'o}n, J.~Verduijn, G.~Lansbergen, G.~Tettamanzi, S.~Rogge, and
  B.~Koiller, ``Heterointerface effects on the charging energy of the shallow
  {D}- ground state in silicon: Role of dielectric mismatch,'' {\em Physical
  Review B}, vol.~82, no.~7, p.~075317, 2010.

\bibitem{kohn1955theory}
W.~Kohn and J.~Luttinger, ``Theory of donor states in silicon,'' {\em Physical
  Review}, vol.~98, no.~4, p.~915, 1955.

\bibitem{ramdas1981spectroscopy}
A.~Ramdas and S.~Rodriguez, ``Spectroscopy of the solid-state analogues of the
  hydrogen atom: donors and acceptors in semiconductors,'' {\em Reports on
  Progress in Physics}, vol.~44, no.~12, p.~1297, 1981.

\bibitem{rahman2009orbital}
R.~Rahman, G.~Lansbergen, S.~H. Park, J.~Verduijn, G.~Klimeck, S.~Rogge, and
  L.~C. Hollenberg, ``Orbital stark effect and quantum confinement transition
  of donors in silicon,'' {\em Physical Review B}, vol.~80, no.~16, p.~165314,
  2009.

\bibitem{szabo2012modern}
A.~Szabo and N.~S. Ostlund, {\em Modern quantum chemistry: introduction to
  advanced electronic structure theory}.
\newblock Courier Corporation, 2012.

\bibitem{slater1954wave}
J.~Slater and G.~Koster, ``Wave functions for impurity levels,'' {\em Phy.
  Rev}, vol.~94, p.~1498, 1954.

\bibitem{rahman2007high}
R.~Rahman, C.~J. Wellard, F.~R. Bradbury, M.~Prada, J.~H. Cole, G.~Klimeck, and
  L.~C. Hollenberg, ``High precision quantum control of single donor spins in
  silicon,'' {\em Physical Review Letters}, vol.~99, no.~3, p.~036403, 2007.

\bibitem{yokota2011treecode}
R.~Yokota and L.~A. Barba, ``Treecode and fast multipole method for {N}-body
  simulation with {CUDA},'' {\em GPU Computing Gems Emerald Edition},
  pp.~113--132, 2011.

\bibitem{nielsen2012many}
E.~Nielsen, R.~Rahman, and R.~P. Muller, ``A many-electron tight binding method
  for the analysis of quantum dot systems,'' {\em Journal of Applied Physics},
  vol.~112, no.~11, p.~114304, 2012.

\bibitem{salfi2014spatially}
J.~Salfi, J.~Mol, R.~Rahman, G.~Klimeck, M.~Simmons, L.~Hollenberg, and
  S.~Rogge, ``Spatially resolving valley quantum interference of a donor in
  silicon,'' {\em Nature Materials}, vol.~13, no.~6, pp.~605--610, 2014.

\bibitem{lansbergen2008gate}
G.~Lansbergen, R.~Rahman, C.~Wellard, I.~Woo, J.~Caro, N.~Collaert,
  S.~Biesemans, G.~Klimeck, L.~Hollenberg, and S.~Rogge, ``Gate-induced
  quantum-confinement transition of a single dopant atom in a silicon
  {FinFET},'' {\em Nature Physics}, vol.~4, no.~8, pp.~656--661, 2008.

\bibitem{taniguchi1976d}
M.~Taniguchi and S.~Narita, ``D- state in silicon,'' {\em Solid State
  Communications}, vol.~20, no.~2, pp.~131--133, 1976.

\bibitem{boykin2004valence}
T.~B. Boykin, G.~Klimeck, and F.~Oyafuso, ``Valence band effective-mass
  expressions in the sp3d5s* empirical tight-binding model applied to a {S}i
  and {G}e parametrization,'' {\em Physical Review B}, vol.~69, no.~11,
  p.~115201, 2004.

\bibitem{calderon2006quantum}
M.~Calderon, B.~Koiller, X.~Hu, and S.~D. Sarma, ``Quantum control of donor
  electrons at the {S}i-{S}i{O}2 interface,'' {\em Physical Review Letters},
  vol.~96, no.~9, p.~096802, 2006.

\bibitem{calderon2009quantum}
M.~J. Calder{\'o}n, A.~Saraiva, B.~Koiller, and S.~Das~Sarma, ``Quantum control
  and manipulation of donor electrons in {S}i-based quantum computing,'' {\em
  Journal of Applied Physics}, vol.~105, no.~12, p.~122410, 2009.

\bibitem{pica2016surface}
G.~Pica, B.~W. Lovett, R.~Bhatt, T.~Schenkel, and S.~Lyon, ``Surface code
  architecture for donors and dots in silicon with imprecise and nonuniform
  qubit couplings,'' {\em Physical Review B}, vol.~93, no.~3, p.~035306, 2016.

\bibitem{voisin2015spatially}
B.~Voisin, J.~Salfi, J.~Bocquel, R.~Rahman, and S.~Rogge, ``Spatially resolved
  resonant tunneling on single atoms in silicon,'' {\em Journal of Physics:
  Condensed Matter}, vol.~27, no.~15, p.~154203, 2015.

\bibitem{salfi2017robust}
J.~Salfi, B.~Voisin, A.~Tankasala, J.~Bocquel, M.~Usman, M.~Simmons, H.~LCL,
  R.~Rahman, and S.~Rogge, ``Observation of valley filtering in {H}eisenberg
  exchange interactions in silicon,'' {\em In preparation}, 2017.

\bibitem{voisin2014control}
B.~Voisin, M.~Cobian, X.~Jehl, M.~Vinet, Y.-M. Niquet, C.~Delerue,
  S.~De~Franceschi, and M.~Sanquer, ``Control of the ionization state of three
  single donor atoms in silicon,'' {\em Physical Review B}, vol.~89, no.~16,
  p.~161404, 2014.

\end{thebibliography}

\end{document}